# Selecting Best Software Reliability Growth Models: A Social Spider Algorithm based Approach

Najla Akram AL-Saati, PhD
Assist. Professor, Software Engineering Dept.
College of Computer Sciences & Mathematics,
University of Mosul, Iraq

Marrwa Abd-AlKareem Alabajee
Assist. Lecturer, Software Engineering Dept.
College of Computer Sciences & Mathematics,
University of Mosul, Iraq

## ABSTRACT
Software Reliability is considered to be an essential part of software systems; it involves measuring the system's probability of having failures; therefore, it is strongly related to Software Quality. Software Reliability Growth Models are used to indicate the expected number of failures encountered after the software has been completed, it is also an indicator of the software readiness to be delivered. This paper presents a study of selecting the best Software Reliability Growth Model according to the dataset at hand. Several Comparison Criteria are used to yield a ranking methodology to be used in pointing out best models. The Social Spider Algorithm (SSA), one of the newly introduced Swarm Intelligent Algorithms, is used for estimating the parameters of the SRGMs for two datasets. Results indicate that the use of SSA was efficient in assisting the process of criteria weighting to find the optimal model and the best overall ranking of employed models.

## 1. INTRODUCTION
Software nowadays can found in all aspects of life, in all scientific, commercial and industrial sectors. It is simply made of a group of code lines that links a specific input(s) into some desirable output(s) carrying out a certain task as defined by the user's requirements. Software, being human written, can very likely contain problems or faults that can lead to an overall system failure. Such failures in software have a direct impact on the reliability and dependability from the user point of view [1]. For such reasons there was a necessity to yield high quality software projects that can function correctly with on-time performance satisfying the given requirements [2]. A software project is defined "as a set of activities with a starting date, specific goals and conditions, defined responsibilities, a budget, a planning, a fixed end date, and multiple parties involved"[3]. The main issue in developing faultless software is reliability, reliable software projects can be expensive and time consuming. Furthermore, the reliability of software has to be calculated to be used in planning test resources throughout the development of software [2][4]. In general, Reliability can be defined as "the probability for failure-free operation of a program for a specified time under a specified set of operating conditions". Software reliability has a direct impact on software quality, and it can be viewed as a key attribute to quality[5]. Assessing software reliability can be done using software reliability growth models (SRGMs). SRGMs offer quantifiable statistics necessary for improving the software reliability of products, software engineers can also benefit from SRGMs in quantifying levels of defect, rates of failure and reliability through the coding and testing phases [2][6]. Various SRGMs have been proposed since 1970 in the literature, yet none of them satisfies all datasets. As Lyu has observed that no universally acceptable model is found that can be trustworthy of giving precise results for all circumstances; every single model embraces some benefits and yet some drawbacks. The selection of the best model for any dataset relays essentially on software requirements [1][7][8]. Swarm intelligence (SI) is a branch of Artificial Intelligence entirely inspired by the social behavior of organisms living and interacting in the interior of large groups of independent individuals. Such behavior can be observed in flocks of birds, Bats, Fireflies, schools of fish, colonies of ants, and even human social behavior. The observed behaviors of swarms can be used for allowing groups of individuals to achieve processes that cannot be done by each single individual by itself [9][10]. Recently, authors are employing SI to obtain feasible solutions for complex optimization problems and in software reliability optimization [11].

In this work, the Weighted Criteria technique proposed by Anjum [7] is applied with the aid of the Social Spider Algorithm (SSA) rather than Least Square and Maximum Likelihood Estimation. SSA is used in the course of estimating the parameters of SRGMs, in order to enhance the performance of criteria weighting to rank the SRGMs according to the best. The Weighted Criteria technique is carried out here with 10 different criteria instead of only 7 to increase efficiency of results.

## 2. LITERATURE REVIEW
Many methods have been proposed in the literature to find a way for selecting the best fit model, such as:

Stringfellow and Andrews, (2002) applied various SRGMs iteratively in system testing; these models were fitted to weekly cumulative failure data. They were used to estimate the expected residual number of failures after software release. When an SRGM passes the proposed criteria, then it is selected to make release decision[12]. In the same year, Kharchenko et al. proposed to choice SRGMs based on the analysis of assumptions and compatibility of input and output parameters, where an assumptions matrix was developed for such choice depending on the features of software engineering and testing processes [13].In 2006, Sheta employed Particle Swarm Optimization (PSO) in estimating the parameters of some of SRGMs such as the exponential model, power model and S-Shaped models[14]. In addition, Garg et al. in 2010 suggested a method based on matrix operations based on performance analysis of SRGMs. They used seven comparison criteria to rank various SRGMs. The result was a ranking of SRGMs based on Permanent value [15]. Also in 2010, Sharma et al. presented a deterministic quantitative model based on distance based approach (DBA) and was applied to select and rank SRGMs [16]. Sharma et al. modified the Artificial Bee Colony (ABC) in 2011, yielding the DABC (Dichotomous ABC), by converging to individual optimal point and to compensate the limited amount of search moves of original ABC. They also explored the use of DABC in estimating SRGMs parameters [17].While the work of Shanmugam and Florence in 2012 solved the parameter estimation problem using Ant Colony Algorithm. Results were gained using six typical models [18]. In 2013, Anjum et al. offered a method based on weighted criteria, where a set of twelve comparison criteria were formulated. Case study results showed that the weighted criteria value method gave a very promising performance in SRGMs comparison [7]. Miglani proposed a guide for the selection of best SRGMs in 2014; the technique was tested on various datasets. The model recommended on the basis of proposed technique has proved to be better in comparison with other recommendations [19]. Sheta and Abdel-Raouf in 2016 investigated the possibility of using the Grey Wolf Optimization (GWO) in the estimation of the SRGM's parameters aiming at minimizing the difference between the estimated and the actual number of failures of the software system [20].

## 3. DEBUGGING PROCESS

Debugging is the process of detecting software faults and correcting them; Saxena et al. divided the process of debugging into the following two types [21] as shown in the following sections.

### 3.1 Perfect Debugging

Perfect debugging involves the correction of faults with certainties responsible for software failures without introducing new faults. Previously introduced software reliability models adopt the fact of perfecting the fault removal process. Jelinski and Moranda presume that the software failure rate is proportional to the number of residual bugs, where each bug owns a constant failure rate impact [22]. Furthermore, the number of bugs drops by one subsequent to each failure designating a flawless elimination of bugs causing the failure. Next are some of the perfect debugging NHPP SRGMs [2]:

1. Goel-Okumoto Model (Goel-O.):

   $$m(t) = a(1 - e^{-bt}) \quad \ldots \ldots \ldots \ldots \ldots \ldots \ldots \ldots (1)$$

   $a > 0, b > 0$

2. Generalized Goel Model (G.Goel):

   $$m(t) = a(1 - e^{-bt^c}) \quad \ldots \ldots \ldots \ldots \ldots \ldots (2)$$

   $a > 0, b > 0, c > 0$

3. Gompert Growth Curve Model (Gompert):

   $$m(t) = ak^{-bt} \quad \ldots \ldots \ldots \ldots \ldots \ldots \ldots \ldots (3)$$

   $a > 0, 0 < b < 1, 0 < k < 1$

4. Inflected S-Shaped Model (Inf.S.):

   $$m(t) = a * \frac{1 - \exp[-bt]}{1 + \beta * \exp[-bt]} \quad \ldots \ldots \ldots \ldots (4)$$

   $a > 0, b > 0, \beta > 0$

5. Logistic Growth Curve Model (Log.Gro.):

   $$m(t) = \frac{a}{1 + k * \exp[-bt]} \quad \ldots \ldots \ldots \ldots \ldots (5)$$

   $a > 0, b > 0, k > 0$

6. Musa-Okumoto Model (Musa-O.)

   $$m(t) = a * ln(1 + bt) \quad \ldots \ldots \ldots \ldots \ldots \ldots (6)$$

   $a > 0, b > 0$

7. Yamada Delayed S-Shaped Model (Y. Del.):

   $$m(t) = a(1 - (1 + bt) * \exp[-bt]) \ldots \ldots (7)$$

   $a > 0, b > 0$

8. Modified Duane Model (Modi-D.):

   $$m(t) = a\left[1 - \left(\frac{b}{b+t}\right)^c\right] \quad \ldots \ldots \ldots \ldots \ldots (8)$$

   $a > 0, b > 0, c > 0$

9. Pham Zhang IFD Model (P-Z-IFD):

   $$m(t) = a - a * \exp[-bt] * (1 + (b + d) * t + bdt^2)$$

   $\ldots \ldots \ldots \ldots (9) \quad a > 0, b > 0, d > 0$

### 3.2 Imperfect Debugging

Imperfect debugging was introduced after noticing that Perfect Debugging is an unrealistic assumption, this is mainly because of the human element involved in software debugging. Each time a new fault is introduced in the correction process and, for some reason, was detected but not removed with certainty, the debugging is called Imperfect Debugging. Below are samples of the Imperfect Debugging models [2][21].

1. Yamada Rayleigh Model (Y. Ray.):

   $$m(t) = a\left(1 - \exp\left[-r\alpha\left(1 - \exp\left[-\frac{\beta t^2}{2}\right]\right)\right]\right) \quad \ldots \ldots$$

   $(10) \, a > 0, r > 0, \alpha > 0, \beta > 0$

2. Yamada Imperfect Debugging Model 1 (Y. M1):

   $$m(t) = a * b * \left(\frac{\exp[\alpha t] - \exp[-bt]}{\alpha + b}\right) \quad \ldots \ldots \ldots \ldots (11)$$

   $a > 0, b > 0, \alpha > 0$

3. Yamada Imperfect Debugging Model 2 (Y. M2):

   $$m(t) = a * (1 - \exp[-bt]) * \left(1 - \frac{\alpha}{b}\right) + \alpha a t \quad \ldots \ldots$$

   $(12) \, a > 0, b > 0, \alpha > 0$

4. Yamada Exponential Model (Y. Exp.):

   $$m(t) = a * (1 - \exp[-\alpha r(1 - \exp[-\beta t])]) \quad \ldots \ldots$$

   $(13) \, a > 0, b > 0, \alpha > 0, \beta > 0$

5. Pham Nordmann Zhang (P–N–Z) model (P-N-Z):

   $$m(t) = \frac{a * (1 - \exp[-bt]) * \left(1 - \frac{\alpha}{b}\right) + \alpha a t}{1 + \beta * \exp[-bt]} \quad \ldots \ldots \ldots \ldots (14)$$

   $a > 0, b > 0, \alpha > 0, \beta > 0$

6. Pham–Zhang Model (P–Z) model:

   $$m(t) = \frac{1}{(1 + \beta * \exp[-bt])}\left((c + a) * (1 - \exp[-bt]) - \frac{ab}{b - \alpha} * (\exp[-\alpha t] - \exp[-bt])\right) \ldots \ldots \ldots \ldots \ldots (15)$$

   $a > 0, b > 0, c > 0, \alpha > 0, \beta > 0$

7. Zhang-Teng-Pham Model (Z-T-P):

   $$m(t) = \frac{a}{p - \beta} * \left(1 - \frac{(1 + \alpha) * \exp[-bt]}{1 + \alpha * \exp[-bt]}\right)^{\frac{c}{b}(p - \beta)} \ldots \ldots \ldots \ldots (16)$$

   $a > 0, b > 0, c > 0, p > 0, \alpha > 0, \beta \geq 0$

## 4. SOCIAL SPIDER ALGORITHM (SSA)

SSA is a newly presented swarm algorithms, it was developed by Yua and Lia [23] for solving global numerical optimization problems. It is built on the bases of the social spiders' behavior to work out solutions for optimization problems. SSA was designed to handle continuous unconstrained problems. This is usually done by formulating the search space of the problem as a hyper-dimensional spider web, where each spider on the web has a specific position; this position denotes a feasible solution to the optimization problem. Artificial spiders in SSA have the ability to move without obstruction on the web, each time a spider changes its position it produces a vibration that is propagated over the web. Here the web functions as a transmission media of the vibrations produced when the spiders move [23]. The following subsections will introduce a more detailed specification of SSA.

### 4.1 Spider

SSA depends largely on the primary functioning agents known as the artificial spiders. Artificial spiders are placed on the web when the algorithm starts. Assuming that(t) is the current iteration index and $f(x)$ is the objective function, each spider (s) in the population is called the $i^{th}$ spider, and it holds two attributes: position $p_i(t)$ and fitness $f(p_i(t))$ for the current position. Each spider owns a memory to store the previous attributes as well as several attributes used to direct the spider to search for the global optimum. Such attributes are [23][24]:

1. The target vibration of (s) in the previous iteration.

2. The number of iterations since (s) has last changed its target vibration.

3. The previous movement that (s) do it in the iteration.

4. The dimension mask that (s) used it to direct the movement in the previous iteration.

### 4.2 Vibration

SSA is recognized by its main vibration feature, the variation is generated and spread across the web each time a spider makes a move to a new position, other spiders on the web will all get that vibration. Spiders in a population are allowed to share their personal information to generate a collective social knowledge of the solution space. Vibrations are recognized using the source position (P) and the source intensity (I), the value of P depends on the search space of the problem, while the I value is limited in the range of [0,+∞) and can be calculated using the fitness value of the position $f(p)$ using Eq.17 [25].

$$I = log\left(\frac{1}{f(P)-C} + 1\right) \quad \ldots\ldots\ldots\ldots\ldots\ldots (17)$$

where

I: is the source intensity, $f(p)$: is the fitness value of (p),

C: is a small constant.

After generating the vibration, it can be propagated across the web; other spiders in the population just receive partial information of the vibration due to the consideration of vibration attenuation in the design of the SSA. The vibration attenuation process is shown in Eq.18 [25]:

$$I^d = I * \exp\left(-\frac{d}{\bar{\sigma}*r_a}\right) \quad \ldots\ldots\ldots\ldots\ldots\ldots (18)$$

where

- $I^d$: is the attenuated intensity after being propagating for distance(d),

- d: is the distance between spiders a and b, calculated using Manhattan distance,

- $\bar{\sigma}$: is the mean of the standard deviation of the population's positions over all dimensions, $\Upsilon_a$: is used for controlling the attenuation rate of the vibration intensity, $\Upsilon_a \in (0,+\infty)$.

The larger $\Upsilon_a$ is the weaker the attenuation of the vibration.

### 4.3 Search in SSA

In order to conduct a search-for-solution procedure in SSA, first the parameters for the algorithm must be set as well as the definition of the fitness function and solution space of the optimization problem. Then a random generation of the initial population of artificial spiders with their positions is performed. An iteration is started following the next steps [25][23]:

**Step1:** Fitness Evaluation: At the start of any iteration, a re-evaluation of the fitness values is performed for each spider on different positions on the web. This evaluation is carried out once for every spider during each iteration.

**Step2:** Vibration Generation: each spider generates a new vibration at its current position using Eq.17. This vibration, after that, is propagated over the web by Eq.18 and is expected by all other spiders. Hence, each spider in the population will receive vibrations by the size of the population |pop|, each spider will choose the one with the largest attenuated vibration intensity $V^{best}$ from |pop|, and then compare it with $V^{tar}$ (the target vibration), if $V^{best}$ is greater, then it is saved as the new $V^{tar}$. When there is no change in the target vibration, then the spider's inactive degree is increased by (1), otherwise this degree is reset to (0).

**Step3:** Mask Changing: in this step a random walk is prepared towards $V^{tar}$, the dimension mask (m) is used to guide the movement. Each spider holds a dimension mask (m), which is a (0-1) binary vector of length D (the dimension of the optimization problem). Throughout the iterations, the spider has a probability of (1- $p_c^{d_{in}}$) to change its mask, where $p_c \in (0,1)$ is user-controlled, and $d_{in}$ is the inactive degree of the spider. If a decision is made to change the mask, then each bit of the mask can be assigned (1) with ($p_m$) probability, and assigned (0) with (1-$p_m$). This probability is user-controlled in the range of (0,1). Bits of a mask are changed independently and don't have any correlation with previous masks. When all bits are (0), one randomly chosen bit of the mask is flipped to (1). Correspondingly, if all bits are (1), one random bit is changed to (0).

**Step4:** Random Walk: after conducting step3, a new following position ($p_s^{fo}$) is generated based on the mask for spider (s). The value of the i$^{th}$ dimension for ($p_{s,i}^{fo}$) is created according to Eq.19.

$$p_{s,i}^{fo} = \begin{cases} p_{s,i}^{tar} & m_{s,i} = 0 \\ p_{s,i}^{r} & m_{s,i} = 1 \end{cases} \quad \ldots\ldots\ldots\ldots\ldots\ldots (19)$$

where

$p_s^{fo}$: a new following position.

r: is a random integer value generated in [1,|pop|],

$m_{s,i}$: is the i<sup>th</sup> dimension of the dimension mask (m) for spider (s),

r: is generated independently for two different dimensions with $m_{s,i}= 1$.

Spider (s) conducts a random walk to the new position using Eq. 20.

$$p_s(t + 1) = p_s + (p_s − p_s(t − 1)) * r + (p_s^{fo} − p_s)⊙R, \quad ........(20)$$

where,

⊙: is element-wise multiplication, R: is a vector of random float-point numbers generated from 0 to 1 uniformly.

Before following ($p_s^{fo}$), spider (s) moves along its previous direction according to the previous iteration. The distance along this direction is a random portion of the previous movement. After that, s approaches ($p_s^{fo}$) along each dimension with random factors generated in (0, 1). This factor is independently generated for different dimensions. After performing this random walk, s stores its movement in the current iteration for the next iteration.

**Step5:** Constraint-Handling: During the previous step, one spider or more may move out of the web. This leads to a violation of the constraints for the optimization problem. Thus, to implement the constraint-handling scheme Eq.21 must be used.

$$p_{s,i}(t + 1) = \begin{cases} (\overline{x_i} − p_{s,i}) * r \text{ if } p_{s,i}(t + 1) > \overline{x_i} \\ (p_{s,i} − \underline{x_i}) * r \text{ if } p_{s,i}(t + 1) > \underline{x_i} \end{cases} \quad ......... (21)$$

where

$\overline{x_i}$: is the upper bound of the search space,

$\underline{x_i}$: is the lower bound of the search space,

r: is a random floating point number generated between (0,1).

When the stopping criterion is met, the iteration is terminated with the best solution for the optimization problem.

## 5. COMPARISON CRITERIA

To study the efficiency of software reliability growth models, an evaluation of the model can be done relying on its capability of reproducing the perceived behavior for the software, and to expect the future behavior of the software from the detected failure data. Thus a number of comparison criteria are suggested in order to carry out a comparison among different proposed models. Comparison criteria are described as follows, where *k* represents the sample size of the data set, and *p* is the number of parameters [16][7]:

1. Bias: describes the sum of the difference between the estimated and the actual data curve as shown in Eq.22.

$$Bias = \frac{\sum_{i=1}^{k}(m(t_i) − m_i)}{k} \quad ................ (22)$$

2. Mean Square Error (MSE): is the deviation between the predicted values and the actual observations as illustrated in Eq.23.

$$MSE = \frac{\sum_{i=1}^{k}(m_i − m(t_i))^2}{k−p} \quad ................ (23)$$

3. Mean Absolute Error (MAE): is the same as MSE, but here the absolute values are used as in Eq.24.

$$MAE = \frac{\sum_{i=1}^{k}|m_i − m(t_i)|}{k−p} \quad ................ (24)$$

4. Mean Error of Prediction (MEOP): is the sum of the absolute value of the difference between the actual data and the estimated curve, this is given in Eq.25.

$$MEOP = \frac{\sum_{i=1}^{k}|m(t_i) − m_i|}{k−p+1} \quad ........... (25)$$

5. Accuracy of Estimation (AE): is the difference between the estimated numbers of all errors with the actual number of all detected errors. Where $M_a$ and (a) are the actual and estimated cumulative number of detected errors after the test, respectively, then Eq.26 shows the formula.

$$AE = \left|\frac{M_a − a}{M_a}\right| \quad ................ (26)$$

6. Noise: is defined as in Eq. 27.

$$Noise = \sum_{i=1}^{k}\left|\frac{\lambda(t_i) − \lambda(t_{i−1})}{\lambda(t_{i−1})}\right| \quad ........... (27)$$

7. Predictive-Ratio Risk (PRR): shows the distance of model estimates from the actual data against the model estimate. It can be formulated as in Eq.28.

$$PRR = \sum_{i=1}^{k} \frac{m(t_i) − m_i}{m(t_i)} \quad ................ (28)$$

8. Variance: is the standard deviation of the prediction bias, it is defined as in Eq. 29

$$Variance = \sqrt{\frac{\sum_{i=1}^{k}(m_i − m(t_i) − Bias)^2}{k−1}} \quad ...(29)$$

9. Root Mean Square Prediction Error (RMSPE): measures the closeness with which the model predicts the observation as given in Eq. 30.

$$RMSPE = \sqrt{Variance^2 + Bias^2} \quad ... (30)$$

10. $R_{sq}$: is a measure of how successful the fit is in explaining the variation of the data. Eq. 31 shows the measure.

$$R_{sq} = 1 − \frac{\sum_{i=1}^{k}(m_i − m(t_i))^2}{\sum_{i=1}^{k}(m_i − \sum_{j=1}^{k}\frac{m_j}{n})^2} \quad ............ (31)$$

11. Sum of Squared Errors (SSE): is formulated as in Eq. 32.

$$SSE = \sum_{i=1}^{k}(m_i − m(t_i))^2 \quad ............. (32)$$

12. Theil Statistic (TS): is the average deviation percentage over all periods with regard to the actual values. The closer TS to zero, the better the prediction capability of the model. It is illustrated in Eq. 33

$$TS = 100 * \sqrt{\frac{\sum_{i=1}^{k}(m_i − m(t_i))^2}{\sum_{i=1}^{k} m_i^2}} \% ... (33)$$

## 6. THE RANKING METHODOLOGY

Considering a multi-attributes decision problem, the formulation of the objective and constraint functions that occur when using a mathematical programming model can be simplified by adopting the approach presented in [7]. This methodology can be used to develop a deterministic quantitative model based on weighted mean, aimed at finding a rank for the software reliability models. To apply this methodology, a matrix is used to denote the value of criteria for each model. Anjum et al. describe the procedure steps as follows [7].-

**Step1:** Constructing the Criteria Value Matrix:

19

A matrix is constructed, where each element $a_{ij}$ is the value of $j^{th}$ criteria of $i^{th}$ model. Assuming that (n) is the number of SRGMs and (m) are the criteria, then this matrix can be given below as:

$$\text{Criteria value matrix} = \begin{bmatrix} a_{11} & a_{12} & \cdots & a_{1m} \\ a_{21} & a_{22} & \cdots & a_{2m} \\ \vdots & \vdots & \cdots & \vdots \\ a_{n1} & a_{n2} & \cdots & a_{nm} \\ (Amin)_1 & (Amin)_2 & \cdots & (Amin)_m \\ (Amax)_1 & (Amax)_2 & \cdots & (Amax)_m \end{bmatrix}$$

where

$(Amax)_j$ = Maximum value of $j^{th}$ criteria,

$(Amin)_j$ = Minimum value of $j^{th}$ criteria,

$a_{ij}$ = Value of $j^{th}$ criteria of $i^{th}$ model.

**Step 2:** Calculating Criteria Rating:

Different software reliability models have different criterion ratings, therefore the criteria rating matrix differs from model to model, it can be determined as:

When smaller value of the criterion represents fitting well to the actual data (Eq.34).

$$X_{ij} = \frac{(Amax)_j - a_{ij}}{(Amax)_j - (Amin)_j} \quad \ldots \ldots \ldots \ldots (34)$$

When bigger value of the criterion represents fitting well to the actual data (Eq.35)

$$X_{ij} = \frac{a_{ij} - (Amin)_j}{(Amax)_j - (Amin)_j} \quad \ldots \ldots \ldots \ldots (35)$$

where

i = (1, 2, 3, ….n) and j = (1, 2, 3, …..m)

**Step3:** Formation of Weight Matrix

Calculating the weight of the criteria can be performed as in Eq. 36.

$$W_{ij} = 1 - X_{ij}, \quad \ldots \ldots \ldots \ldots (36)$$

where

$X_{ij}$: is the matrix of criteria rating.

$$\text{Weight Matrix} = \begin{bmatrix} W_{11} & W_{12} & \cdots & W_{1m} \\ W_{21} & W_{22} & \cdots & W_{2m} \\ \vdots & \vdots & \cdots & \vdots \\ W_{n1} & W_{n2} & \cdots & W_{nm} \end{bmatrix}$$

**Step4:** Building the Weighted Criteria Value Matrix

Weighted criteria value can be computed using Eq.37.

$$A_{ij} = W_{ij} * a_{ij} \quad \ldots \ldots \ldots \ldots (37)$$

$$\text{Weighted Criteria Value Matrix} = \begin{bmatrix} A_{11} & A_{12} & \cdots & A_{1m} \\ A_{21} & A_{22} & \cdots & A_{2m} \\ \vdots & \vdots & \cdots & \vdots \\ A_{n1} & A_{n2} & \cdots & A_{nm} \end{bmatrix}$$

**Step5:** Computation of Model's Permanent Value

The Permanent value of a model is the weighted mean value of all criteria, Eq. 38.

$$Z_i = \frac{\sum_{i=1}^{m} A_{ij}}{\sum_{i=1}^{m} W_{ij}} \quad \ldots \ldots \ldots \ldots (38)$$

where

i: is (1, 2, 3, … n)

Ranking of modelsis carried out using the permanent value of a model; here smaller permanent value of model reflects good rank, opposing to bigger values. Therefore, a comparison is performed among all permanent values to provide ranking for each model.

# 7. EXPERIMENTAL RESULTS
## 7.1 Parameter Settings

Four parameters are used in the Social Spider Algorithm; these are:

1. Population Size (pop): it determines the individual diversity and influences the convergence speed.
2. Attenuation Rate ($r_a$): it defines the rate of vibration attenuation while propagating over the spider web.
3. Mask Changing Probabilities ($p_c$) and ($p_m$): These two cooperate to determine the dimension mask of each spider in the population. $p_c$ controls the probability of changing spider's dimension mask in the random walk step, and $p_m$ defines the probability of each value in a dimension mask to be one.

In this work, these parameters are set as follows :( pop=40, $r_a$=1, $p_c$=0.7, $p_m$=0.1)

## 7.2 Case Study1

The Dataset employed in this work(Phase2 dataset) is used as Dataset1by [26], it includes the number of faults detected in 21 week of testing, and the cumulative number of faults since the starting of test is recorded for each week. Phase2 data observes 416hours per week of testing [26].Sixteen SRGMs are selected for investigation in this study; first the Social Spider Algorithm is applied for estimating the parameters of these models, Table 1 shows the values of estimated parameters for Dataset1 using SSA.

**Table 1. Parameter Estimation using SSA (Dataset1)**

| | Model | Parameter Values |
|---|---|---|
| 1 | Goel-O. | a=4.5457*10^(3), b=4.6771*10(-4) |
| 2 | G.Goel | a=51.5350, b=0.0099, c=1.7236 |
| 3 | Gompert | a=53.6928, b=0.8669, k =0.0179 |
| 4 | Inf. S. | a=46.3662, b=0.2580, β=14.8860 |
| 5 | Log.Gro. | a=45.8508, b=0.2741, k= 19.9806 |
| 6 | Musa-O. | a=2.7617*10^(3), b=7.7032*10(-4) |
| 7 | Y. Del. | a=60.8072, b=0.1232 |
| 8 | Modi-D. | a=1.1528*10^(3), b=196.8210, c=0.3779 |
| 9 | P-Z-IFD | a=58.0815, b=0.1420, d=0.0091 |
| 10 | Y. Ray. | a=108.7096, α = 0.5833, β=0.0095 |
| 11 | Y. M1 | a=651.9094, b=0.0030, α =0.0128 |
| 12 | Y. M2 | a=2.5130, b=0.3456, α =0.9875 |
| 13 | Y. Exp. | a=749.2519, α= 7.1016, β=3.9344*10(-4) |
| 14 | P-N-Z | a=45.0971, b=0.2605, α = 0.0012, β=14.6798 |
| 15 | P-Z | a=46.7686, b=0.2064, C=7.2006, α =0.1156, β=2.6478 |
| 16 | Z-T-P | a=581.8986, b=0.1460, c=0.0353, α =0.3392, β=11.2635 |

After estimating the parameters of the selected SRGMs, the Weighted Criteria technique is engaged using the ten of the comparison criteria explained in section5 (MSE, MAE, MEOP, AE, Noise, RMSPE, SSE,TS, PRR, $R_{sq}$) rather than the seven criteria employed by [7] to rank the different SRGMs, results are shown in Table 2. Table 3 shows the Model's Permanent value and ranking using Dataset1.The fifth row in Table 3 shows that the Logistic Growth Model is the best model suitable for Dataset1. Figure 1 illustrates the actual and estimated failures for the Logistic Growth Model using SSA Values of actual and estimated failures are very close together indicating the optimality of the model.

## 7.3 Case Study2

The second dataset used in this paper is (DS1) used by [27], this dataset is used in this work as Dataset2.47.65 CPU hours were spent in 19 weeks, and 328 software faults have been found and removed [27]. SSA was also used for parameter estimation as Table 4 shows the values of estimated parameters for dataset2. After that, the weighted criteria method is applied also using the same ten comparison criteria used in Case Study1 to rank the different SRGMs. Table 5 demonstrates the values for the selected models. In Table 6, the Model permanent values and ranking are given for Dataset2. Table 6 shows that Z-T-P model is the best model that suitable for Dataset2. Figure2 Depicts the actual and estimated failures of Z-T-P model using SSA. Here again the actual and estimated failure values are very close to each other signifying that the first ranked model is optimal.

**Table 2. Criteria Values of SRGMs (Dataset1)**

| Model/Criteria | MSE | MAE | MEOP | AE | Noise | RMSPE | SSE | TS | PRR | Rsq |
|---|---|---|---|---|---|---|---|---|---|---|
| Goel-O. | 6.6637 | 2.2751 | 2.1614 | 0.0233 | 0.0094 | 2.6789 | 126.6111 | 9.1599 | 1.0198 | 0.9693 |
| G.Goel | 3.0991 | 1.7045 | 1.6148 | 0.0233 | 2.1581 | 1.6712 | 55.7831 | 6.0800 | -6.4582 | 0.9865 |
| Gompert | 2.1885 | 1.4235 | 1.3486 | 0.0233 | 10.7874 | 1.4058 | 39.3935 | 5.1094 | -1.4299 | 0.9904 |
| Inf. S. | 2.0470 | 1.2351 | 1.1701 | 0 | 5.5569 | 1.3837 | 36.8456 | 4.9414 | -4.4564 | 0.9911 |
| Log.Gro. | 1.1412 | 0.9049 | 0.8572 | 0 | 2.9727 | 1.1025 | 20.5410 | 3.6895 | -4.4564 | 0.9911 |
| Musa-O. | 6.7084 | 2.2716 | 2.1580 | 0.0233 | 0.0153 | 2.6629 | 127.4587 | 9.1905 | 0.9983 | 0.9691 |
| Y. Del. | 3.4349 | 1.7784 | 1.6894 | 0.0233 | 2.0994 | 1.8222 | 65.2638 | 6.5764 | -7.5830 | 0.9842 |
| Modi-D. | 8.0185 | 2.5304 | 2.3973 | 0 | 0.0600 | 2.7755 | 144.3333 | 9.7800 | 1.1054 | 0.9650 |
| P-Z-IFD | 3.8053 | 1.9004 | 1.8004 | 0.0233 | 3.1817 | 6.7373 | 68.4963 | 1.8865 | -325.9355 | 0.9834 |
| Y. Ray. | 3.6088 | 1.7644 | 1.6715 | 0.0233 | 3.1675 | 1.8391 | 64.9592 | 6.5611 | -11.9533 | 0.9842 |
| Y. M1 | 6.2901 | 2.2754 | 2.1556 | 0.0698 | 0.2049 | 2.4489 | 113.2227 | 8.6621 | 0.9983 | 0.9691 |
| Y. M2 | 5.3514 | 2.0944 | 1.9841 | 0.0930 | 0.6582 | 2.2841 | 96.3260 | 7.9896 | -1.6504 | 0.9766 |
| Y. Exp. | 8.6727 | 2.6004 | 2.4636 | 0 | 0.0634 | 2.8811 | 156.1078 | 10.1711 | 0.3945 | 0.9621 |
| P-N-Z | 2.1437 | 1.3043 | 1.2318 | 0 | 2.6767 | 1.3850 | 36.4426 | 4.9143 | -4.4704 | 0.9912 |
| P-Z | 3.0792 | 1.7359 | 1.6338 | 0.0233 | 2.7651 | 1.7523 | 49.2667 | 5.7139 | -6.7189 | 0.9880 |
| Z-T-P | 4.2792 | 1.9509 | 1.8362 | 0 | 3.4787 | 2.2287 | 68.4666 | 6.7359 | -33.5239 | 0.9834 |

**Table3. Permanent Values and Ranking (Dataset1)**

|  | Model | Sum of weight | Sum of weighted value | Model Value | Rank |
|---|---|---|---|---|---|
| 1 | Goel-O. | 117.8944 | 5.3644 | 21.9772 | 12 |
| 2 | G.Goel | 21.0328 | 3.4581 | 6.0821 | 6 |
| 3 | Gompert | 19.7532 | 3.5080 | 5.6310 | 5 |
| 4 | Inf. S. | 10.2226 | 2.5051 | 4.0808 | 3 |
| 5 | Log. Gro. | 1.8606 | 1.2840 | 1.4490 | 1 |
| 6 | Musa-O. | 119.4365 | 5.3616 | 22.2762 | 13 |
| 7 | Y. Del. | 29.3619 | 3.7429 | 7.8446 | 8 |
| 8 | Modi-D. | 155.7421 | 5.7294 | 27.1832 | 14 |
| 9 | P-Z-IFD | 359.3462 | 5.0692 | 70.8875 | 16 |
| 10 | Y. Ray. | 29.9845 | 3.8666 | 7.7548 | 7 |
| 11 | Y. M1 | 94.0768 | 5.5175 | 17.0506 | 11 |
| 12 | Y. M2 | 67.1463 | 5.4126 | 12.4056 | 10 |
| 13 | Y. Exp. | 182.8969 | 6.0050 | 30.4574 | 15 |
| 14 | P-N-Z | 8.0126 | 2.3270 | 3.4434 | 2 |
| 15 | P-Z | 17.0227 | 3.5384 | 4.8108 | 4 |
| 16 | Z-T-P | 38.1385 | 4.2558 | 8.9616 | 9 |

**Table 4. Parameter Estimation SSA (Dataset2)**

|  | Model | Parameter Values |
|---|---|---|
| 1 | Goel-O. | a=738.9787, b=0.0335 |
| 2 | G.Goel | a=431.3436, b=0.0361, c=1.2780 |
| 3 | Gompert | a=385.9318, b=0.0483, c=0.8487 |
| 4 | Inf. S. | a=381.7563, b=0.1774, β=2.7918 |
| 5 | Log. Gro. | a=347.2247, b=0.2846, k= 10.6625 |
| 6 | Musa-O. | a=640.2760, b=0.0386 |
| 7 | Y. Del. | a=369.8893, b=0.2017 |
| 8 | Modi-D. | a=1.0724*10^(3), b=95.4546, c=2.1796 |
| 9 | P-Z-IFD | a=369.9175, b=0.2018, d=9.3438e-005 |
| 10 | Y. Ray. | a=540.5019, α = 0.9836, β=0.0155 |
| 11 | Y. M1 | a=774.0665, b=0.0315, α=1.7150*10^(-4) |
| 12 | Y. M2 | a=883.6043, b=0.0268, α=2.2544e-005 |
| 13 | Y. Exp. | a=811.3068, (r*α)= 8.6992, β=0.0035 |
| 14 | P-N-Z | a=290.1209, b=0.1511, α=0.0251, β=1.0722 |
| 15 | P-Z | a=211.2503, b=0.1844, c=167.3081, α=40.0734, β=3.0363 |
| 16 | Z-T-P | a=212.5387, b=0.2466, c=0.3025, α=9.8988, β=0.5939 |

**Table 5. Criteria values of SRGMs for Dataset2**

| Model/Criteria | MSE | MAE | MEOP | AE | Noise | RMSPE | SSE | TS | PRR | Rsq |
|---|---|---|---|---|---|---|---|---|---|---|
| Goel-O. | 156.5539 | 11.0704 | 10.4554 | 0.0610 | 0.5930 | 12.5235 | 2661.4 | 5.2771 | 0.6510 | 0.9864 |
| G.Goel | 121.4961 | 9.4312 | 8.8764 | 0.0366 | 1.0999 | 10.5401 | 1943.89 | 4.5100 | -0.3768 | 0.9901 |
| Gompert | 103.2497 | 8.4137 | 7.9188 | 0.0274 | 9.5592 | 9.5848 | 1652.0 | 4.1576 | 0.2290 | 0.9916 |
| Inf. S. | 98.5457 | 8.2750 | 7.7882 | 0.0244 | 2.9237 | 9.3823 | 1576.7 | 4.0618 | -0.1223 | 0.9920 |
| Log. Gro. | 108.1181 | 8.1123 | 7.6351 | 0.0091 | 2.5916 | 9.8495 | 1729.9 | 4.2545 | 0.3640 | 0.9912 |
| Musa-O. | 166.3821 | 11.3840 | 10.7515 | 0.0732 | 0.5048 | 12.8768 | 2828.5 | 5.4402 | 0.6046 | 0.9856 |
| Y. Del. | 190.1881 | 11.2141 | 10.5911 | 0.0091 | 2.3755 | 14.1097 | 3233.20 | 5.8164 | -2.8140 | 0.9835 |
| Modi-D. | 169.6865 | 11.8601 | 11.1625 | 0.0671 | 10.6590 | 12.5391 | 2715.0 | 5.3299 | 0.5855 | 0.9862 |
| P-Z-IFD | 202.3143 | 11.9158 | 11.2149 | 0.0091 | 2.3779 | 14.1303 | 3237 | 5.8198 | -2.8313 | 0.9835 |
| Y.Ray. | 324.2411 | 13.7609 | 12.9514 | 0.0061 | 3.5059 | 20.1835 | 5187.9 | 7.3677 | -6.1168 | 0.9735 |
| Y. M1 | 166.2916 | 11.7372 | 11.0468 | 0.0640 | 0.5540 | 12.3293 | 2660.7 | 5.2763 | 0.5651 | 0.9864 |
| Y. M2 | 171.7253 | 11.7006 | 11.0123 | 0.0762 | 0.4755 | 12.3982 | 2747.6 | 5.3618 | 0.4260 | 0.9860 |
| Y. Exp. | 167.3101 | 11.7926 | 11.0989 | 0.0610 | 0.5826 | 12.3886 | 2677 | 5.2925 | 0.6101 | 0.9863 |
| P-N-Z | 141.8075 | 10.7410 | 10.0697 | 0.0549 | 1.3576 | 10.8955 | 2127.1 | 4.7177 | 0.2888 | 0.9892 |
| P-Z | 112.9091 | 9.4423 | 8.8128 | 0.0244 | 2.0134 | 9.4392 | 1580.7 | 4.0669 | -0.2514 | 0.9919 |
| Z-T-P | 91.4009 | 8.0141 | 7.4798 | 0.0183 | 2.4192 | 8.4909 | 1279.6 | 3.6591 | 0.2446 | 0.9935 |

**Table 6. Permanent Values and Ranking (Dataset2)**

| | Model | Sum of weight | Sum of weighted value | Model value | Rank |
|---|---|---|---|---|---|
| 1 | Goel-O. | 1003.7 | 4.0190 | 249.7447 | 12 |
| 2 | G.Goel | 354.5 | 2.5744 | 137.7159 | 6 |
| 3 | Gompert | 174.8 | 2.6428 | 66.1315 | 3 |
| 4 | Inf. S. | 126.5 | 1.8185 | 69.5552 | 4 |
| 5 | Log. Gro. | 210.7 | 1.6845 | 125.0868 | 5 |
| 6 | Musa-O. | 1195.8 | 4.4016 | 271.6751 | 13 |
| 7 | Y. Del. | 1721.5 | 4.2911 | 401.1755 | 15 |
| 8 | Modi-D. | 1087.8 | 5.4221 | 200.6151 | 8 |
| 9 | P-Z-IFD | 1745.8 | 4.5851 | 380.7546 | 14 |
| 10 | Y. Ray. | 5573.5 | 7.2976 | 763.7493 | 16 |
| 11 | Y. M1 | 1015.5 | 4.2571 | 238.5525 | 9 |
| 12 | Y. M2 | 1113.2 | 4.4760 | 248.7051 | 11 |
| 13 | Y. Exp. | 1034 | 4.2932 | 240.8343 | 10 |
| 14 | P-N-Z | 506.4 | 3.4655 | 146.1123 | 7 |
| 15 | P-Z | 139.2 | 2.2090 | 62.9943 | 2 |
| 16 | Z-T-P | 1.5 | 1.3852 | 1.0565 | 1 |

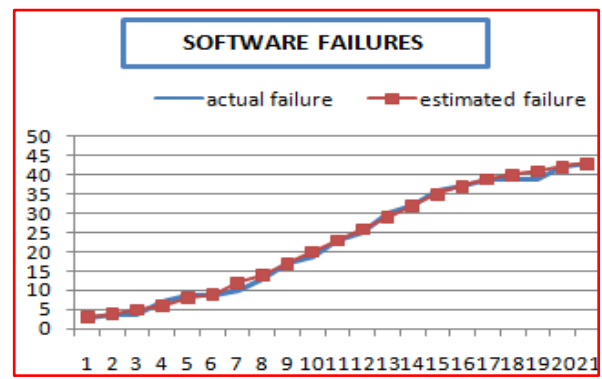

**Fig 1: Actual and estimated failures for the Logistic Growth Model using SSA (Dataset1)**

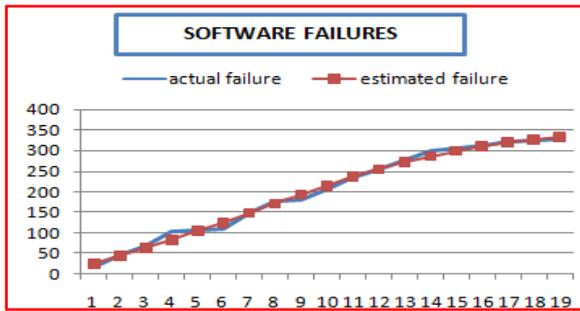

**Fig 2: Actual and estimated failures for Z-T-P model using SSA (Dataset2)**

## 8. CONCLUSIONS

According to the fact that there is no optimal growth model of Software Reliability suitable for all the involved criteria, then a unified criterion is needed to decide the most appropriate model for a given dataset.This work discusses the question of selecting the optimal software reliability growth model by using the weighted matrix method applied on two datasets of failure. The Social Spider Algorithm (SSA) was used for parameter estimation instead of relying on parameter estimated using the Least Square and Maximum Likelihood Estimation.The weighted criteria method was used to determine the overall rank of a model. Results were improved by using SSA. Ranks of Models were provided accordingly for the two selected datasets, and the best ranked models were proven to be the optimal after comparing the actual and estimated failures for these models.